\documentstyle[aps]{revtex} 
\begin{document}
\input epsf
\def\la{\mathrel{\mathpalette\fun <}}
\def\ga{\mathrel{\mathpalette\fun >}}
\def\fun#1#2{\lower3.6pt\vbox{\baselineskip0pt\lineskip.9pt
  \ialign{$\mathsurround=0pt#1\hfil##\hfil$\crcr#2\crcr\sim\crcr}}}
\def\3he{$^3$He}
\def\4he{$^4$He}
\def\6li{$^6$Li}
\def\7li{$^7$Li}
\def\he3{$^3$He}
\def\eg{{\it e.g.}}
\def\ie{{\it i.e.}}
\def\etal{{\it et al.~}}
\def\hii{H\thinspace{$\scriptstyle{\rm II}$}~}
\def\popii{Pop\thinspace{$\scriptstyle{\rm II}$}~}
\def\Yp{Y$_{\rm p}$}
\def\adot{\odot{a}}
\draft
\twocolumn[\hsize\textwidth\columnwidth\hsize\csname
@twocolumnfalse\endcsname

\title{Observational Constraints On Power-Law Cosmologies}
                     
\author{M. Kaplinghat$^1$, G. Steigman$^{1,2}$, I. Tkachev$^{3,4}$, and 
                    T. P. Walker$^{1,2}$}
\address{
{\it $^1$Department of Physics, The Ohio State University, Columbus, OH 
43210\\ 
$^2$Department of Astronomy, The Ohio State University, Columbus, OH 
43210\\}}

\address{
{\it $^3$Department of Physics, Purdue University, West Lafayette, IN 
47907\\
$^4$Institute for Nuclear Research of the Academy of Sciences of Russia, 
Moscow, Russia\\}}

\date{ April 1998}
\maketitle

%

\begin{abstract}

In a class of models designed to solve the cosmological constant problem 
by coupling scalar or tensor classical fields to the space-time curvature,
the universal scale factor grows as a power law in the age, $a \propto 
t^\alpha$, regardless of the matter content or cosmological epoch.  We 
investigate constraints on such ``power-law cosmologies" from the present 
age of the Universe, the magnitude-redshift relation, and from primordial 
nucleosynthesis.  
Constraints from the current age of the Universe and from the high-redshift 
supernovae data require ``large" $\alpha$ ($\approx 1$), while consistency 
with the inferred primordial abundances of deuterium and helium-4 forces 
$\alpha$ to lie in a very narrow range around a lower value ($\approx 0.55$).
Inconsistency between these independent cosmological constraints suggests
that such power-law cosmologies are not viable.

\end{abstract}
\vskip2pc]
%


\section{Introduction}

According to General Relativity all mass/energy gravitates, including 
the energy density of the vacuum.  In modern quantum field theory the 
vacuum is the lowest energy -- but not necessarily the zero energy -- 
state.  From this perspective a cosmological constant ($\Lambda$) may
be associated with the vacuum energy density, $\rho_{\rm vac} = 
\Lambda/8\pi G \equiv \Omega_\Lambda \rho_{\rm c}$, where $\rho_{\rm c}
\equiv 3{{\rm H}_0}^2/8\pi G \sim 10^{-48} {\rm GeV}^4$ is the critical
density.  Although some recent data favor a non-zero value of $\Lambda$ 
\cite{Ia}, observations do limit 
$\Omega_\Lambda \la 1$\cite{lambda}  corresponding
to a vacuum energy density that is very 
small when compared to that expected from physics at the Planck scale ($\sim
10^{19}$ GeV).    This is because although 
we may wish to set $\Lambda = 0$ in the Einstein equations, quantum 
fluctuations in the fields present in the Universe can establish a non-zero 
vacuum energy and, hence, a non-zero effective cosmological constant.  We 
may associate the vacuum energy density with an energy  
scale $M$ which might be the scale associated with the spontaneous symmetry 
breaking from one vacuum state to another, $\rho_{\rm vac}\sim M^4$.  In 
some sense the only ``natural" 
scale in cosmology is the Planck scale, $M \sim 10^{19}$~GeV.  In this 
case the observations require that the present vacuum energy density is 
some 120 orders of magnitude smaller than its ``natural" value.  The smallness 
of $\rho_{\rm vac}$ is a key problem in modern
cosmology: the ``$\Lambda$" or ``cosmological constant problem''\cite{swein}.


One class of attempts to solve the $\Lambda$-problem considers the 
evolution of classical fields which are coupled to the curvature of 
the space-time background in such a way that their contribution to 
the energy density self-adjusts to cancel the vacuum energy\cite{regulate}.
Although 
the dynamical framework in these approaches is well defined, the addition 
of the special fields is unmotivated but for solving the cosmological 
constant problem.    The common result of these approaches is that the 
vacuum energy 
may be nearly cancelled and the expansion of the Universe is governed by 
the uncompensated vacuum energy density.  In such models the expansion 
is a power-law in time, independent of the matter content or cosmological 
epoch (see Ford, ref\cite{regulate}).  That is, in such models 
the scale factor varies according to $a(t) 
\propto t^\alpha$, where $\alpha$ is determined solely by the parameters 
of the model and can be anywhere in the range $0\le\alpha\le \infty$. In
addition, there are  models designed to solve other cosmological fine-tuning 
problems (\eg, flatness \cite{Allen}) which also result in power-law 
cosmologies.

In this {\sl Letter} we explore
the constraints on $\alpha$ from the age-expansion rate data, from the
magnitude-redshift relation of type Ia supernovae (SN~Ia) at redshifts 
0.4 -- 0.8, and from the requirement that primordial nucleosynthesis 
produce deuterium and helium-4 in abundances consistent with those 
inferred from observational data. 

\section{Constraints from The Age/Expansion Rate}

In power-law cosmologies the scale factor $a(t)$, the redshift $z$, and
the CMB temperature $T(t)$ are related to their present values (labelled
by the subscript ``0") by

\begin{equation}
a/a_{0} = 1/(1 + z) = T_{0}/\beta T = (t/t_{0})^{\alpha}.
\end{equation}
where $\beta$ accounts for any non-adiabatic expansion due to entropy 
production
(\eg, in standard cosmology $\beta = 1$ for $T < m_e$ and $\beta =
(11/4)^{1/3}$ for $T > m_e$ accounting for the heating due to
$e^{\pm}$ annihilation  assuming instantaneous annihilation at $T=m_e$).
All models (all choices of $\alpha$) are ``normalized" by requiring
that they have have the current temperature, $T_{0} = 2.728$K\cite{COBE}, at 
present ($t_0$).  For the present age of the Universe
we adopt $t_{0} = 14 \pm 2$~Gyr \cite{age}; we explore the small effect on 
our constraints of
this choice for $t_0$.  The Hubble parameter, $H = \dot{a}/a$ provides
a measure of the expansion rate.  For power-law cosmologies, $Ht = 
\alpha$, so that at present $H_{0}t_{0} = \alpha$.  If we adopt a 
central estimate and allow for a generous uncertainty in the Hubble
parameter $H_{0} = 70 \pm 10$~km~s$^{-1}$~Mpc$^{-1}$\cite{Hnaught}, we limit $\alpha$:
\begin{equation}
\alpha = H_{0}t_{0} = 1.0 \pm 0.2.
\end{equation} 
Consistency with the present age of the Universe suggests that 
$\alpha \ga 0.6$.  In order for $\alpha$  to be as small as $ 0.5$,  
$H_{0} \sim 50$~km~s$^{-1}$~Mpc$^{-1}$ and $t_{0}$ $\sim 10$~Gyr.

\section{Constraints From The Magnitude-Redshift Relation}

The expansion of the Universe in power-law cosmologies is completely 
described by the Hubble parameter and the deceleration parameter.  In 
these models the deceleration parameter is

\begin{equation}
q(t)= - H^{-2}(\ddot{a}/a) = q_{0} = \frac{1}{\alpha} - 1,
\end{equation} 
so that for $\alpha \ga 1/2$, $q_{0} \la 1$.  The larger $\alpha$, 
the smaller $q_{0}$ and, vice-versa.  As $\alpha$ (= $H_{0}t_{0}$) 
increases from 1/2 to 1, $q_{0}$ decreases from 1 to 0; negative 
values of $q_{0}$ require $\alpha > 1$. 

For spatially flat power-law cosmologies the luminosity distance and/or 
angular-diameter distance redshift relations assume a very simple form.  
The luminosity distance (in units of the Hubble distance $c/H_{0}$) as 
a function of redshift is

\begin{equation}
d_{\rm L}(z) = q_{0}^{-1}[(1 + z) - (1 + z)^{(1 - q_{0})}].
\end{equation} 
Note that for $\alpha = 1/2$ ($q_{0} = 1$), $d_{\rm L}(z) = z$ 
while for $\alpha = 1$ ($q_{0} = 0$), $d_{\rm L}(z) = z(1 + z)$.
Recently two independent groups\cite{Ia} have been accumulating observations of 
possible ``standard candles", SN~Ia, at relatively high redshifts 
($z \sim 0.4 - 0.8$).  The difference in apparent magnitudes of 
objects with the same intrinsic luminosity but at different redshifts 
provides a valuable, classical cosmological test. 
The figure of merit for power-law cosmologies is the expected difference in 
apparent magnitudes, 
$\Delta m \equiv 5log[d_{\rm L}(z_{2})/d_{\rm L}(z_{1})]$, 
for $z_{1} = 0.4$ and $z_{2} = 0.8$ as a function of $\alpha$.  As 
$\alpha$ increases from 1/2 to 1, $\Delta m$ increases from 1.5 
(magnitudes) to 1.8.  For comparison, the recent discovery of a SN~Ia at 
$z = 0.83$ by 
Perlmutter \etal \cite{Ia} suggests that $\Delta m \approx 2.0 \pm 0.2$, 
favoring a small (or even negative) $q_{0}$, corresponding to a ``large" 
value of $\alpha ~\ga 1$.  These data seem to exclude $\Delta m \la 
1.6$ which corresponds to $\alpha \la 0.6$.  Much more data have been 
accumulated since this published report and if they confirm this result, 
viable power-law cosmologies will be restricted to those with relatively 
large values of $\alpha$ ($\ga 0.6$).  Primordial nucleosythesis  
provides a powerful constraint on such models.

\section{Constraints from Primordial Nucleosynthesis}

As we've seen above, observations of the recent Universe ($z \la 1$)
favor values of $\alpha$ close to unity.  Such power-law models are 
a disaster for primordial nucleosynthesis.  For example, suppose that 
$\alpha = 1$, as results in some
$\Lambda$-regulating 
models\cite{regulate} or as proposed in Allen\cite{Allen}, and ask how 
old was the Universe when 
nucleosynthesis began at a temperature of order 80 keV?  From equation 
(1) the answer is 45 years!  At this stage any neutrons have long since 
decayed and there can be no nucleosynthesis\footnote{We verify in our 
numerical 
results presented below that pp or pep reactions are inadequate to 
compensate for the absence of neutrons.}.  This simple example helps to 
focus the physical origin of the BBN challenge to power-law cosmologies.  
Unless a suitable early time-temperature relation exists, neither helium-4 
nor deuterium will be produced primordially in amounts comparable to those
inferred from the observational data.  

In our discussion we concentrate on the abundances of $^4$He and D for 
the following reasons.  Observations reveal that the most metal-poor 
stars and/or \hii regions have a {\it minimum}, non-zero abundance of 
$^4$He; for this primordial mass fraction we adopt the generous range 
0.22 $\leq$ Y$_{\rm P} \leq$ 0.26\cite{helium} .  Any viable cosmological 
model 
must account for this much $^4$He.  Similarly, the observation of significant 
abundances of deuterium requires primordial production\cite{deut} .  Here, 
too, we adopt a generous range $1\times 10^{-5} 
\leq$ D/H $\leq 2 \times 10^{-4}$\cite{tpw} .  Any model producing too much 
or too little 
deuterium is excluded.  Note that if/when we identify a viable power-law 
model consistent with these D and \4he constraints, we do check the 
consistency of the predicted abundances of $^3$He 
and $^7$Li.

To understand BBN in power-law cosmologies it is helpful to briefly
review primordial nucleosynthesis in the standard model (SBBN).  At high 
temperature ($\ga 1$~MeV) charged-current weak interactions among neutrons, 
protons, electrons, positrons, and neutrinos maintain neutron-proton 
equilibrium: $n/p = \exp (-Q/T)$ (where Q = 1.29 MeV is the neutron-proton 
mass difference).  In SBBN the weak interaction rates interconverting 
neutrons and protons become slower than the universal expansion rate for
T $\la 1$ MeV (when the Universe is of order 1~s old) and the $n/p$ ratio 
``freezes out", decreasing only very slowly due to out-of-equilibrium 
weak interactions and free neutron decay (with a lifetime of 887~s).  All
the while neutrons and protons are colliding to form deuterium which is
rapidly photodissociated by the cosmic background photons (gamma rays at
this epoch).  The very low abundance of D removes the platform for building
heavier nuclei; nucleosynthesis is delayed by this ``photodissociation
bottleneck''.  When the temperature drops below $\sim 80$~keV (the Universe
is $\sim 3$ minutes old) the deuterium bottleneck is broken and nuclear
reactions quickly burn the remaining free neutrons ($n/p \approx 1/7$) 
into $^4$He (Y$_{\rm P} \approx 0.25$), leaving behind trace amounts 
of D, $^3$He, and $^7$Li\cite{bbn}.  
If 
the light elements are to be properly synthesized during BBN, the above 
scenario must be mimicked in a viable power-law cosmology.

First, let us consider the photodissociation bottleneck and free neutron 
decay.  To ensure that {\it some} primordial nucleosynthesis will occur
neutrons must have not decayed before the deuterium bottleneck is broken.
Thus, we require that $t~\la 887~$s when T $\approx 80$~keV.  This leads
immediately to a constraint on $\alpha$ (which depends only logarithmically 
on our choice of $t_{0} = 14$~Gyr); from equation (1), $\alpha$ $\la 0.58$.

{\it Power-law models which succeed in having {\bf any} BBN are in conflict 
with the constraints from the present age/Hubble parameter and the SN~Ia 
magnitude-redshift relations discussed above.}

To explore BBN in power-law cosmologies in more detail it is important 
to understand how the time-temperature relation in these models changes 
with $\alpha$.  In Figure 1 the time-temperature relation is shown for 
several choices of $\alpha$.

\begin{figure}[Fig1]
\centering
\leavevmode\epsfxsize=8.7cm \epsfbox{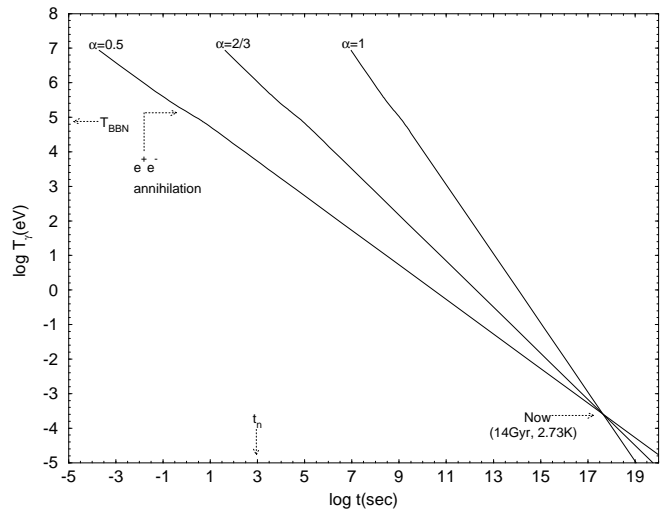}\\
\
\caption[Fig1]{\label{fig1} The age-temperature relation for three 
power-law cosmologies 
($\alpha$ = 1/2, 2/3, and  1).  Time is measured in seconds and temperature 
in eV.  T$_{\rm BBN} \approx 80$~keV is the temperature at which 
nucleosynthesis begins; t$_{\rm n} = 887~s$ is the neutron lifetime; 
t$_{0} = 14$~Gyr is the present age of the Universe (where T$_0 = 2.73$K).  
Note the slight ``kink'' due to entropy production at $e^{\pm}$ annihilation }
\end{figure}

The larger $\alpha$, the faster the Universe 
expands.  For example, for a fixed time early on (say, $t = t_{n} =$ 
887~s) the higher $\alpha$, the higher the temperature.  Similarly, if we 
fix on a definite temperature in the early Universe, the higher $\alpha$ 
the older the Universe.  This may seem counterintuitive 
because, although the Universe (with higher $\alpha$) is indeed expanding 
faster, it is {\it younger} at a fixed temperature for {\it lower} $\alpha$ 
since all models (choices of $\alpha$) are constrained to have T$_{0} = $
2.7~K at 14~Gyr.  Thus the requirement that BBN occur before the 
free neutrons decay bounds $\alpha$ from above. 

There also exists a {\it lower} bound to $\alpha$ since 
for low $\alpha$ (young Universe) the Universe will have too little 
time for nuclear reactions to build up any significant abundances of 
the light elements.  As $\alpha$ decreases, the weak interactions decouple
at higher temperatures and neutrons have less time to decay thus leading to 
a larger neutron fraction at the time of nucleosynthesis.  Provided 
nuclear reactions are efficient, this increased neutron fraction results in
more \4he.  However, once $\alpha$ becomes sufficiently small, nuclear 
reactions
become inefficient and no nucleosynthesis occurs.  For small enough $\alpha$,
\4he should decrease with decreasing $\alpha$.  The critical $\alpha$ 
delineating these regimes depends
on the nucleon density since a young age can be compensated 
by having a higher nucleon density leading to faster nuclear reaction 
rates.  We have, therefore, explored BBN numerically for a wide range of 
choices of $\alpha$ and of $\eta$, the universal nucleon-to-photon ratio 
($\eta \equiv n_{\rm N}/n_{\gamma}$; $\eta_{10} \equiv 10^{10}\eta$).

In Figure 2 the evolution of the neutron-to-proton ratio as a function 
of temperature is shown for several choices of $\alpha$. 

\begin{figure}[Fig2]
\centering
\leavevmode\epsfxsize=8.7cm \epsfbox{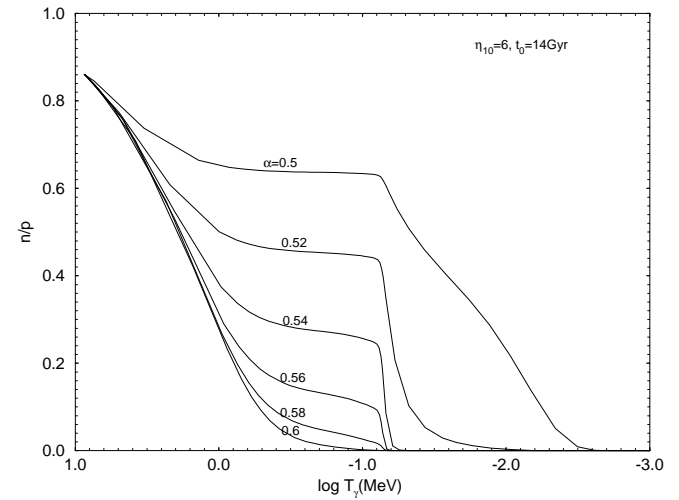}\\
\
\caption[Fig2]{\label{fig2} The neutron-to-proton ratio as a function of 
temperature 
for several choices of $\alpha$.  T$_{\rm BBN} \approx 80$~keV is 
the temperature at which nucleosynthesis begins.   }
\end{figure}

For T$~\ga 
80$~keV, the decline in $n/p$ reflects neutron decay; the larger $\alpha$, 
the older the Universe (for fixed T), and the more neutrons have decayed.  
The precipitous decline in $n/p$ for T$~\la 80$~keV is due mainly to 
nuclear reactions incorporating free neutrons in the light nuclides.
As expected from our semi-analytic argument above, if $\alpha$ is too
large, too few neutrons remain when BBN can begin.  Note that the smaller
$\alpha$ the larger the freeze-out abundance of neutrons (at $\sim$ 1 MeV)
 and the smaller
the effects of neutron decay.  If $\alpha$ is too
small, nuclear reactions are inefficient at forming heavier nuclei and the
decline after 80 keV is not as severe.

In Figures 3 and 4 we concentrate on the {\it interesting} range of
$\alpha$ and show the predicted abundances of $^4$He (Fig. 3) and of
D (Fig. 4) for a variety of $\eta$ values covering more than an order
of magnitude.  In Figure 5 we show iso-abundance contours in the 
$\eta - \alpha$ plane corresponding to the observed ranges for D and \4he 
adopted above.  As anticipated, there is a very narrow range of $\alpha$ 
(0.552 -- 0.557) which results in an acceptable yield of primordial 
\4he and D for $2 \la \eta_{10} \la 15$.  If we impose a further constraint
from \7li \cite{li}, the upper bound on $\eta_{10}$ is reduced to 12.

\begin{figure}[Fig3]
\centering
\leavevmode\epsfxsize=8.7cm \epsfbox{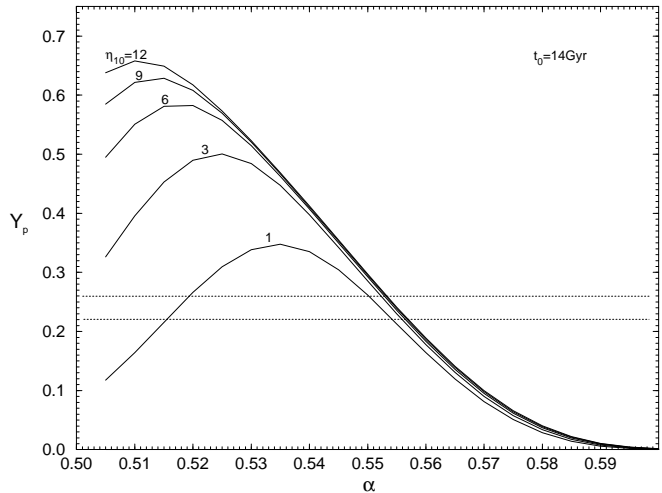}\\
\
\caption[Fig3]{\label{fig3} The primordial $^4$He mass fraction, Y$_{\rm P}$, 
as a function of $\alpha$ for several choices of the nucleon-to-photon 
ratio, $\eta_{10} \equiv 10^{10}n_{\rm N}/n_{\gamma}$.  The horizontal 
dotted lines delimit the adopted range of the primordial abundance.}
\end{figure}

\begin{figure}[Fig4]
\centering
\leavevmode\epsfxsize=8.7cm \epsfbox{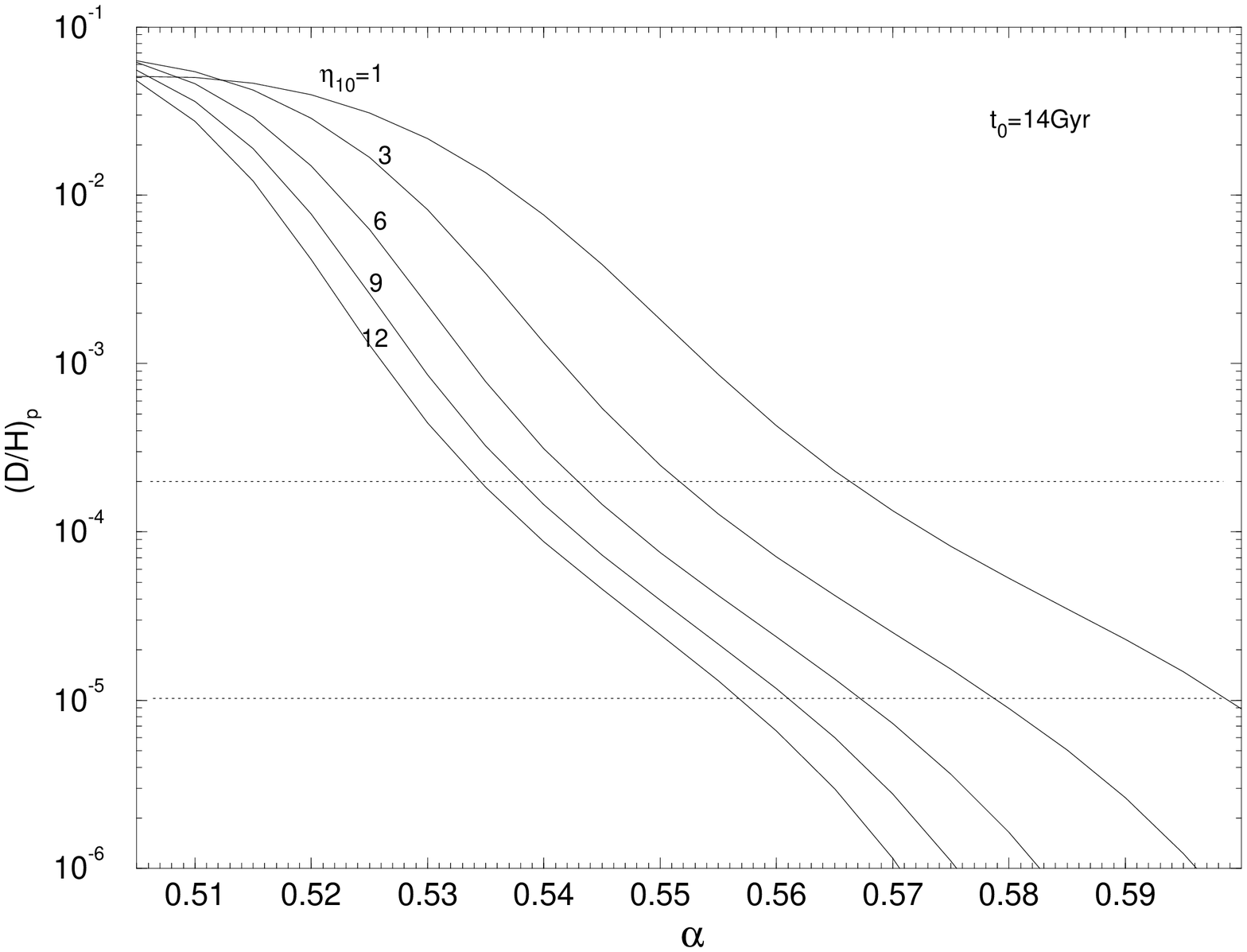}\\
\
\caption[Fig4]{\label{fig4} As for Figure 3 for the primordial 
deuterium abundance, D/H.}
\end{figure}

\begin{figure}[Fig5]
\centering
\leavevmode\epsfxsize=8.7cm \epsfbox{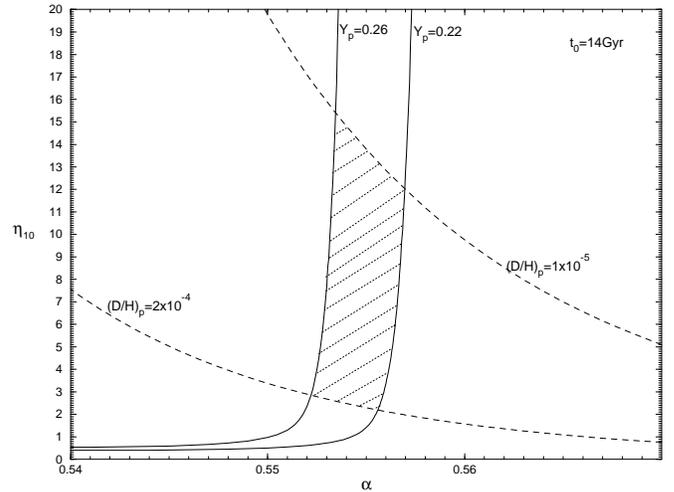}\\
\
\caption[Fig5]{\label{fig5} Iso-abundance contours in the 
$\eta_{10}$ -- $\alpha$ plane
corresponding to the observational bounds on D and \4he. }
\end{figure}

A simple heuristic argument will serve to expose the physical 
origin of this concordant range for $\alpha$.  For these values of 
$\alpha$, the power-law cosmologies are evolving very similarly to 
the standard model for 1~MeV $\ga$ T $~\ga$ 30~keV.  For $\alpha$ 
in this very narrow range the ages at fixed temperature are, within 
factors of order unity, the same as those in SBBN (\ie, for $\alpha 
\sim 0.55$, the temperature is around 1 MeV when the Universe is 1 
second old and the temperature is around 100 keV when it is 1 minute 
old, ensuring that weak freeze out and the onset of nucleosynthesis 
work in concert as in SBBN).  It is worth noting that this range for 
$\alpha$ is insensitive (logarithmically) to our choice of 14~Gyr for 
the present age of the Universe. 

As Figure 3 reveals, there is a second, lower range of values of $\alpha$ 
which, depending on $\eta$, might yield acceptable primordial helium.  
Although such models have very high neutron fractions when BBN commences 
(see Fig. 2), these models are so young the time for complete BBN is 
insufficient (unless the nucleon density is sufficiently high).  This is 
shown dramatically in Figure 4 where the very high deuterium abundances 
reflect the incomplete burning to $^4$He.  Note that the primordial yields
in this low $\alpha$ limit are very sensitive to $\eta$  since the yields 
are set by a competition between expansion and nuclear reaction rates.  
However, these lower values of $\alpha$ do not provide viable power law 
models from the BBN perspective since they cannot simultaneously produce 
the correct abundances of \4he and D.  In the limit of low $\alpha$, D is 
always overproduced relative to \4he since the Coulomb barriers involved 
in \4he production inhibit the burning of D to \4he.

There is one caveat we should note concerning our constraint on
power-law cosmologies from BBN.  Basically, we have seen that BBN 
requires that a potentially successful model have a time-temperature
relation which crosses the lower left-hand region in Figure 1 where
t $\sim 887 s$ (the neutron lifetime) when T = 80 keV.  This
requirement has permitted the exclusion of larger values of $\alpha$.
If, however, entropy is released 
during or after BBN, a
successful time-temperature relation may have existed during BBN
for values of $\alpha$ larger than those allowed in the absence of
such entropy production.  For example, entropy may be released through
the decay of a massive particle such as those associated with the
moduli fields of supersymmetry (supergravity) \cite{ss} theory.   Given 
the extra free parameters associated with this possibility (amount and 
timing of the entropy release; \eg, mass and lifetime of the decaying 
massive particle(s)), we have chosen to not explore here this option for 
avoiding our BBN constraints on power-law cosmologies.  

\section{Discussion}

Can the evolution of the Universe -- from very early epochs to the present 
-- be described by a simple power law relation between the age and the scale 
factor (temperature)?  In standard cosmology the early Universe is radiation
dominated (RD) and the expansion is a power law with $\alpha_{\rm RD} = 
1/2$.  But, in standard cosmology the Universe switched from RD to matter
dominated (MD) at a redshift between 10$^{3}$ and 10$^{4}$.  Thereafter 
the Universe expanded (for a while at least) according to a power law with 
a different power: $\alpha_{\rm MD} = 2/3$.  If the present Universe has 
a low density (compared to the critical density) and lacks a significant
cosmological constant, it is ``curvature" dominated (CD) and its expansion 
may be well approximated by a power law with $\alpha_{\rm CD} = 1$.  Thus
in standard cosmology, although power law expansion may provide a good 
description for some epochs, there is no single power which can describe 
the entire evolution from, for example, BBN to the present.  The question 
then is, can a ``compromise" $\alpha$ be found which is consistent with 
BBN as well as with observations of the present/recent Universe?  

We have explored this question and answered it in the negative.  The present
age/expansion rate (Hubble parameter) constraint $\alpha = H_{0}t_{0} = 
1.0 \pm 0.2$ and the SN~Ia magnitude-redshift relation require $\alpha 
\approx 1$ (or, $\alpha$ $\ga$ 0.6), while production of primordial helium
and deuterium force $\alpha$ to be smaller.  The extreme sensitivity of
the helium yield to $\alpha$ (see Fig. 3), precludes raising the upper
bound on $\alpha$ from BBN.  Unless the Universe is much younger ($\la 
$~10~Gyr) and/or the Hubble parameter much smaller ($\la~$50~km~s$^{-1}$
Mpc$^{-1}$) than currently believed and the SN~Ia magnitude-redshift
relation plagued by systematic errors, or there was substantial entropy
release after BBN, power law cosmologies are not the 
solution to the cosmological constant problem.

\acknowledgments

We thank R. Scherrer for helpful discussions and suggestions.
This work was supported at Ohio State by DOE grant DE-AC02-76ER01545.

\end{document}